\documentstyle[preprint,prb,aps]{revtex}
 \begin{document}
 \draft
 \title{Does spin-orbit coupling play a role in metal-nonmetal
transition  in two-dimensional  systems?}

 \author{Guang-Hong Chen, M. E. Raikh, and  Yong-Shi Wu}
 \address{Department of Physics, University of Utah, 
 Salt Lake City, Utah  84112}
 \maketitle
 \begin{abstract}
We propose an experiment, which would allow to 
pinpoint the role of spin-orbit coupling in the 
metal-nonmetal transition observed in a number 
of two-dimensional systems at low densities. 
Namely, we demonstrate that in a parallel magnetic 
field the interplay between the spin-orbit 
coupling and the Zeeman splitting leads to a 
characteristic anisotropy of resistivity {\em with 
respect to the direction of the  in-plane magnetic 
field}. Though our analytic calculation is done in 
the deeply insulating regime, the anisotropy is 
expected to persist far beyond that regime.

\end{abstract}

 
\narrowtext

In a recent paper\cite{papada99} an interesting 
experimental observation was reported. It was 
demonstrated that the period of beats of the 
Shubnikov-de Haas oscillations in a two-dimensional
hole system is strongly correlated with the 
zero-magnetic-field temperature dependence of the 
resistivity. The beats of the Shubnikov-de Haas 
oscillations have their origin in the splitting of 
the spin subbands in a zero magnetic field,\cite{luo,Das}. 
The authors of Ref. \onlinecite{papada99} were able 
to tune the zero-field splitting by changing the gate 
voltage. They observed that, while in the absence of the 
subband splitting, the zero-magnetic-field resistivity 
was temperature-independent below $T=0.7$K , a pronounced 
rise (by $5$ percent) in resistivity with temperature
emerged in the interval $0.2K < T < 0.7K$ at the maximal 
subband splitting, indicating a metallic-like behavior. 
This close correlation suggests that it is a mechanism 
causing the spin subband splitting that plays an 
important role in the crossover from the metallic-like 
to the insulating-like temperature dependence of 
resistivity with decreasing carrier density (the 
metal-nonmetal transition). This transition by now 
has been experimentally observed in a number of 
different two-dimensional electron\cite{kra,krav,pop,han} 
and hole\cite{lam,col,pap,hane,sim} systems. By challenging the
commonly accepted concepts, it
has attracted a lot of theoretical interest 
and attempts to identify the underlying mechanism.
Possible relevance of zero-field splitting
to the transition was first conjectured in 
Ref. \onlinecite{pud97}. The evidence  presented in 
Ref. \onlinecite{papada99}  about the importance of 
the subband splitting for metallic-like behavior 
of resistivity is further supported by the very 
recent data reported in Ref. \onlinecite{Yaish99}. 



Another important feature of the metal-nonmetal transition, 
which might also provide a clue for the understanding of 
its origin, is that the metallic phase is destroyed 
by a relatively weak {\em parallel} magnetic 
field\cite{sim,simonian97,prinz,kravchenko98,simonian98,mertes99,okamoto98}. 
At the same time, no quenching of the metallic phase 
in a parallel magnetic field was observed in SiGe hole 
gas\cite{senz99}, in which the strain, caused by the 
lattice mismatch, splits the light and heavy holes. 

As far as the theory is concerned,  the role 
of the parallel magnetic field was previously accounted 
for exclusively through the Zeeman energy, which either 
alters the exchange interactions (and, thus, electron-ion 
binding energy\cite{klap,alt}) or suppresses the liquid 
phase,\cite{he} or affects the transmittancy of the point 
contact between the phase-coherent regions\cite{meir}. 

 It is appealing to combine the observations
\cite{papada99,Yaish99} of the  subband splitting 
in zero field and the results
\cite{sim,simonian97,prinz,kravchenko98,simonian98,mertes99,okamoto98} 
in a parallel magnetic field within a single picture. The 
spin-orbit (SO) coupling appears to be a promising candidate for such a 
unifying mechanism. Indeed, on one hand, it is known to lead 
to spin subband splitting. On the other hand, a parallel magnetic 
field, though not affecting the orbital in-plane motion, destroys 
the SO coupling and, thus, suppresses the intersubband transitions.
Possible  importance of these transitions was emphasized in 
Ref. \onlinecite{Yaish99}. Their suppression with increasing
 magnetic field 
is caused by the fact that the corresponding  subband
 wave functions
become orthogonal for {\em all} wave vectors.

At the present moment there is no consensus  in the 
literature about the role of the SO coupling. Several authors
\cite{skvortsov97,dmitriev,lyanda'} have explored the role of 
the SO coupling as a possible source of the metallic-like 
behavior, by considering  noninteracting two-dimensional system and
 including the SO terms 
into the calculation of the weak-localization corrections. At the same
time,
the majority of theoretical works
\cite{klap,alt,he,meir,dobro,belitz,chak1,phillips,cast,si,das,neil,chak2}, 
stimulated by the experimental observation of the transition, 
disregarded the SO coupling.

To pinpoint the role of the SO coupling in the metal-nonmetal 
transition,  it seems important to find a {\em qualitative} 
effect which exists only in the presence of the SO coupling. 
Such an effect is proposed in the present paper. We show that 
an interplay between the SO coupling and the
 Zeeman splitting gives rise to 
a characteristic {\em anisotropy} of resistivity with respect 
to the {\em direction} of the parallel magnetic field. 
Obviously, the Zeeman splitting alone cannot induce any 
anisotropy. To demonstrate the effect, we consider the deeply 
insulating regime, where the physical picture of transport
is transparent. 

We choose the simplest form for the spin-orbit Hamiltonian
\cite{rashba,rashba'}
\begin{equation}
\label{hso}
\hat{H}_{SO}=\alpha{\bf k}\cdot({\bf\bbox{ \sigma}}\times {\bf \hat{z}}).
\end{equation}
Here  $\alpha$ is the SO coupling constant,  ${\bf k}$ is the wave vector,
${\bf \hat{z}}$ is the unit vector normal to the  2D  plane,
  $\bbox{\sigma}=(\sigma_{1},\sigma_{2},\sigma_{3})$  are the Pauli
matrices. In the presence of the parallel megnetic field, the single
particle Hamiltonian can be written as
\begin{equation}
\label{ham}
\hat{H}=\frac{\hbar^2k^2}{2m}+\alpha{\bf k}\cdot({\bf\bbox{ \sigma}}\times
{\bf \hat{z}})+g\mu_B{\bf\bbox{ \sigma}}\cdot{\bf
B}=\left(\begin{array}{cc}\frac{\hbar^2k^2}{2m} & \frac{\Delta_Z}{2}
e^{-i\phi_{\bf B}}-i\alpha k e^{-i\phi_{{\bf k}}}\\ \frac{\Delta_Z}{2}
e^{i\phi_{\bf B}}+i\alpha k e^{i\phi_{\bf k}} &
\frac{\hbar^2k^2}{2m}\end{array}\right),
\end{equation}
where $m$ is the effective mass, $g$ and $\mu_B$ are the g-factor and the
Bohr magneton respectively; $\Delta_Z=2g\mu_B B$ is the Zeeman splitting;
$\phi_{\bf B}$ and $\phi_{\bf k}$ are, correspondingly, the azimuthal
angles of magnetic field ${\bf B}$ (Fig. 1) and the wave vector ${\bf k}$.
The energy spectrum of the Hamiltonian Eq. (\ref{ham}) is given by
\begin{equation}
\label{spectrum}
E_{\pm}({\bf
k})=\frac{\hbar^2k^2}{2m}\pm\frac{1}{2}\sqrt{\Delta_Z^2+4\alpha^2k^2+4\alpha
k \Delta_Z\sin(\phi_{{\bf B}}-\phi_{{\bf k}})}.
\end{equation}
Note that the spectrum is anisotropic only if {\em both} $\Delta_Z$ and
$\alpha$ are nonzero.

The standard procedure for the calculation of
 the hopping conductance is the
following\cite{book}. Denote with  $P_{12}$  the hopping probability
between the localized states $1$ and $2$. The logarithm of $P_{12}$
represents the sum of two terms
\begin{equation}
\label{log}
\ln P_{12}=-\frac{\varepsilon_{12}}{T}-\ln|G({\bf R})|^2,
\end{equation}
where the first term originates from the activation; $\varepsilon_{12}$ is
the activation energy\cite{book} and $T$ is the temperature. The second
term in Eq. (\ref{log}) describes the overlap of the wave functions of the
localized states centered at points ${\bf R}_1$ and ${\bf R}_2$, so that
${\bf R}={\bf R}_1-{\bf R}_2$. In Eq. (\ref{log}) we use the fact that
within the prefactor the overlap integral coincides with the Green
function $G({\bf R})$. For the matrix Hamiltonian Eq. (\ref{ham}), the
Green function is also a matrix
\begin{equation}
\label{green}
\hat{G}({\bf R})=\int\frac{d^2{\bf k}}{(2\pi)^2}\frac{e^{i{\bf k}\cdot{\bf
R}}}{E-\hat{H}({\bf k})}.
\end{equation}
By projecting onto the eigen-space of Hamiltonian Eq. (\ref{ham}), the
above expression can be presented as
\begin{equation}
\label{proj}
\hat{G}({\bf R})=\int\frac{dk k d\phi_{\bf
k}}{(2\pi)^2}e^{ikR\cos(\phi_{\bf k}-\phi_{\bf
R})}\biggl[\frac{\hat{P}_{+}({\bf k})}{E-E_{+}({\bf
k})}+\frac{\hat{P}_{-}({\bf k})}{E-E_{-}({\bf k})}\biggr], 
\end{equation}
where the projection operators $P_{\pm}({\bf k})$ are defined as
\begin{equation}
\label{p+-}
\hat{P}_{+}({\bf k})=\frac{1}{2}\left(\begin{array}{cc}1&
O({\bf k})\\ O^{*}({\bf k}) & 1\end{array}\right),
\hspace{1.2cm}\hat{P}_{-}({\bf k})=1-\hat{P}_{+}({\bf k}),
\end{equation}
where $O^{*}({\bf k})$ is the complex conjugate of $O({\bf k})$, which is defined as
\begin{equation}
\label{off}
O({\bf k})=\frac{\frac{\Delta_Z}{2}\exp(-i\phi_{\bf B})-i\alpha k\exp(-i\phi_{{\bf
k}})}{E_{+}({\bf k})-E_{-}({\bf k})}.
\end{equation}

When the distance $R$ is much larger than the localization radius, $a_0$,
the integral over $\phi_{\bf k}$ is determined by a narrow interval
$|\phi_{\bf k}-\phi_{\bf R}|\sim (kR)^{-1/2}\ll 1$. This allows to replace
$\phi_{\bf k}$ by $\phi_{\bf B}$ in the square brackets and perform the
angular integration. Then we obtain
\begin{equation}
\label{green2}
\hat{G}({\bf R})=\sqrt{\frac{2\pi}{iR}}\int_0^{\infty}\frac{dk
\sqrt{k}}{(2\pi)^2}e^{ikR}\biggl[ \frac{\hat{P}_{+}(k,\phi_{\bf
R})}{E-E_{+}(k,\phi_{\bf R})}+\frac{\hat{P}_{-}(k,\phi_{\bf
R})}{E-E_{-}(k,\phi_{\bf R})}\biggr].
\end{equation}
The next step of the integration is also standard. Namely, for large $R$,
the $\hat{G}({\bf R})$ is determined by the poles of the integrand.
However, in the case under the consideration,
 the equation $E_{\pm}(k)=E$
leads to a fourth-order algebraic equation. To simplify the calculations
we will restrict ourselves to the strongly localized regime $|E|\gg
m\alpha^2/\hbar^2$. In this case the poles can be found by the successive
approximations. In the zero-order approximation, we get the standard
result $k=ik_0$,
where $k_0$ is defined as
\begin{equation}
\label{k0}
k_0=a_0^{-1}=\frac{\sqrt{2m|E|}}{\hbar}.
\end{equation}
 In the first order approximation, we have $k=ik_0+k_1$ where $k_1$ is
given by
\begin{equation}
\label{k1}
k_1=\pm
i\frac{m\alpha}{\hbar^2}\sqrt{\Delta_1^2-1+2i\Delta_1\sin(\phi_{\bf
B}-\phi_{\bf R})},
\end{equation}
where the dimensionless Zeeman splitting $\Delta_1$ is defined as 
\begin{equation}
\label{delta}
\Delta_1=\Delta_Z/2\alpha k_0.
\end{equation}
Within this approximation, the long-distance asymptotics of the Green
function is
\begin{equation}
\label{asym}
\hat{G}(R)\propto e^{-R/a(\phi_{\bf B},\phi_{\bf R})},
\end{equation}
where the decay length is given by
\begin{equation}
\label{decay}
a(\phi_{\bf B}, \phi_{\bf R})^{-1}
=k_0\biggl(1-\frac{m\alpha}
{\hbar^2} Re\sqrt{\Delta_1^2-1+2i\Delta_1\sin(\phi_{\bf B}
-\phi_{\bf R})}\biggr).
\end{equation}
In the last equation it is assumed that the real
part, $Re(...)$, has a positive sign.
Our main observation is that the decay length and, 
concommitantly, the probability of hopping are {\em anisotropic}, 
when the parallel megnetic field and the SO coupling are present
{\em simultaneously}.
By evaluating the real part in Eq. (\ref{decay}) we obtain
\begin{equation}
\label{1/a}
a(\phi_{\bf B}, \phi_{\bf R})^{-1}
=k_0\biggl(1-\frac{m\alpha}{\sqrt{2}\hbar^2k_0}
\sqrt{\Delta_1^2-1+\sqrt{1+\Delta_1^4-2\Delta_1^2
\cos2(\phi_{\bf B}-\phi_{\bf R})}}\biggr).
\end{equation}
To characterize the anisotropy quantitatively, we introduce 
the perpendicular decay length~ $a_{\perp}=
a(\phi_{\bf B}-\phi_{\bf R}=\pm\frac{\pi}{2})$ and the 
parallel decay length $a_{\parallel}=a(\phi_{\bf B}=\phi_{\bf R})$. 
Then a quantitative measure of the anisotropy can be defined as
\begin{equation}
\label{anisotropy}
\frac{a_{\perp}-a_{\parallel}}{a_0}
=\frac{m\alpha}{\hbar^2k_0}f(\Delta_1),
\end{equation}
where the  function $f(x)$ is given  by
\begin{equation}
\label{universal}
f(x)=x-(x^2-1)^{1/2}\theta(x-1),
\end{equation}
where $\theta(x)$ is the step-function. In the strongly localized regime ($\alpha k_0\ll |E|$)  the
anisotropy is weak. The magnetic field dependence of
the anisotropy is shown in Fig. 2. It can be seen that the maximal 
anisotropy corresponds to $\Delta_1=1$ and it vanishes both in  strong
and weak magnetic fields. The theory of hopping transport in the 
systems with
anisotropic localization radius is presented in
Ref. \onlinecite{book}. The principal outcome of this theory is
 that the anisotropy of
the localization radius (and, consequently, the exponential anisotropy
of the hopping probability (\ref{log})) {\em does not} lead to the {\em exponential} anisotropy
of the hopping resistance. In fact, the exponent
 of the resistance is the
same as  for the isotropic hopping with localization radius
$\sqrt{a_{\parallel}a_{\perp}}$. However, the anisotropy in the Green
function manifests itself in the prefactor of the hopping resistance\cite{book}
\begin{equation} 
\label{hopres}
\frac{\rho_{\parallel}-\rho{\perp}}{\rho_{\parallel}+\rho{\perp}}=C\frac{a_{\parallel}-a_{\perp}}{a_{\parallel}+a_{\perp}}\approx C\frac{m\alpha}{2\hbar^2k_0}f(\Delta_1),
\end{equation}
where $C \sim 1$ is the numerical factor, determined by the perturbation theory in
the method of invariants for random bond percolation problem\cite{book}.
The exact value of the constant $C$ depends on the regime of 
hopping (nearest-neighbor or variable-range hopping).
 
The microscopic origin of the SO  Hamiltonaian Eq. (\ref{hso}) 
is the asymmery of the confinement potential. In III-V semiconductor 
quantum wells there  exists another mechanism of the SO coupling, 
which originates from the absence of the inversion symmetry in the 
bulk (the Dresselhaus mechanism\cite{dress}). Within this mechanism
$\hat{H}_{SO}=\beta(\sigma_xk_x-\sigma_yk_y)$ (for [001] growth 
direction). Then the calculation similar to the above leads to 
the following result for the anisotropic decay length
\begin{equation}
\label{anidress}
a(\phi_{\bf B}, \phi_{\bf R})^{-1}
=k_0\biggl(1-\frac{m\beta} {\sqrt{2}\hbar^2k_0}
\sqrt{\Delta_2^2-1+\sqrt{1+\Delta_2^4 
+2\Delta_2^2\cos(\phi_{\bf B}+\phi_{\bf R})}}\biggr),
\end{equation}
where $\Delta_2$ is related to the Zeeman splitting as
\begin{equation}
\label{dell}
\Delta_2=\Delta_Z/2\beta k_0.
\end{equation}
By using Eq. (\ref{anidress}) we get for anisotropy
\begin{equation}
\label{dress}
\frac{a_{\perp}-a_{\parallel}}{a_0}=
-\frac{m\beta}{\hbar^2k_0}f(\Delta_2),
\end{equation}
where the function $f$ is determined by Eq. (\ref{universal}).

In conclusion, we have demonstrated that, due to the SO 
coupling, the rotation of an in-plane magnetic field with 
respect to the direction of current should lead to a 
characteristic angular variation of resistivity with a
period $\pi$. The anisotropy is maximal for intermediate 
magnetic fields and vanishes in the weak and the 
strong-field limits. In the strongly localized regime, 
considered in the present paper, the magnitude of anisotropy 
is small. However, as seen from Eqs. (\ref{anisotropy})
and (\ref{hopres}), the magnitude of anisotropy should 
increase as the Fermi level moves up with increasing  
carrier concentration (since $k_0$ decreases). So the 
resistivity is expected to remain anisotropic, perhaps 
with a modified angular dependence, far beyond the 
deeply insulating regime. For high enough concentrations
the SO coupling (subband splitting) is negligible, so that
the anisotropy should be also weak. If the intersubband
scattering governs the metal-nonmetal transition, then
the resistivity anisotropy should reach maximum around the 
critical density.  
 
Finally, let us discuss two possible complications for the
experimental observation of the anisotropy in  resistivity. 
Both of them stem from the fact that a realistic two-dimensional 
system has a finite thickness. Firstly, with finite thickness, 
even a small deviation of the magnetic field direction 
from the in-plane position would cause a certain anisotropy even
without SO coupling. However, in this case, the anisotropy would
only increase with increasing magnetic field, while the SO-induced
anisotropy should vanish in the strong-field limit.
The second effect of the finite thickness is that it
causes the anisotropy of  the Dresselhaus term with respect to
the crystalline axes.  As it is shown in Ref. \onlinecite{malsh},
the interplay of anisotropic Dresselhaus and isotropic Bychkov-Rashba
terms results in the {\em crystalline} anisotropy of the resistivity
in the weak-localization regime. This effect should be distinguished
from the anisotropy {\em with respect to the direction of current}
predicted in the present paper.
  
{\it Acknowledgements}. The authors are grateful to R. R. Du for 
helpful discussions. M.E.R. acknowledges the support of the NSF 
grant INT-9815194, and Y.S.W. the NSF grant PHY-9601277.  
\newpage

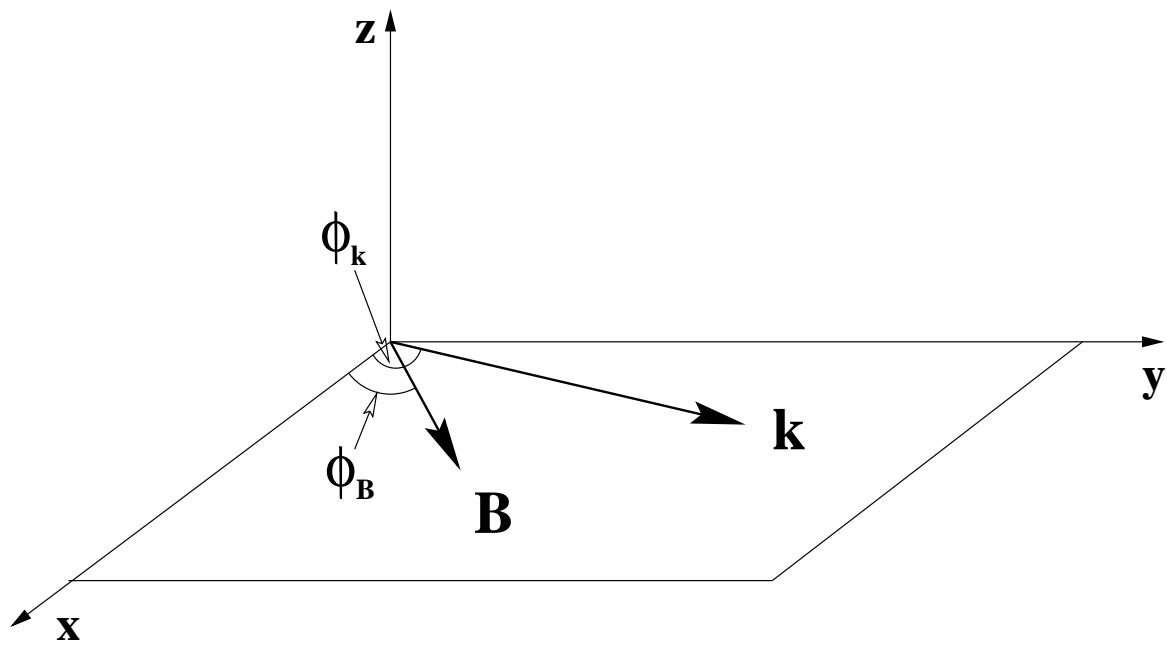
\begin{figure}
\caption{Azimuthal position of the in-plane magnetic field, 
$\phi_{\bf B}$, and of the wave vector of electron, $\phi_{\bf k}$, 
are shown schematically.}
\end{figure}
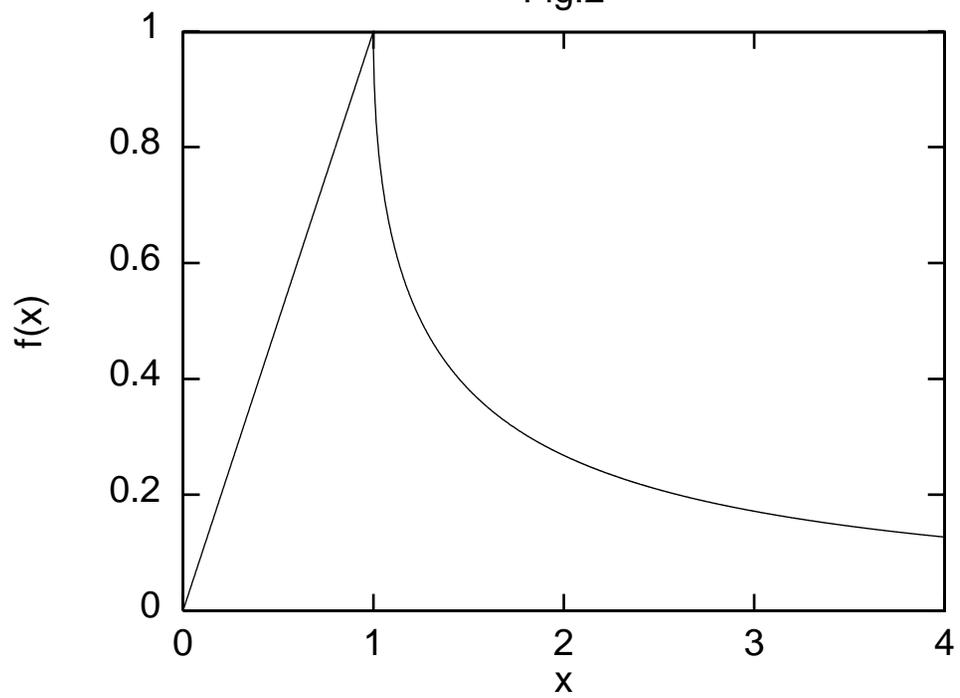
\begin{figure}
\caption{Magnetic field dependence of the anisotropy
of the decay length.}
\end{figure}


\end{document}